\documentclass{PoS}

\title{Probing the MSSM flavor structure \\ with low energy CP violation}
\ShortTitle{Probing the MSSM flavor structure with low energy CP violation}

\author{
\speaker{Wolfgang Altmannshofer}
\thanks{
It is a pleasure to thank the other authors of the work presented here, Andrzej~J.~Buras, Stefania~Gori, Paride~Paradisi and David~Straub for the interesting collaborations. The work of WA is supported by the German Bundesministerium f\"ur Bildung und Forschung under contract 05HT6WOA and the Graduiertenkolleg GRK 1054 of DFG.}
\\
Physik Department, Technische Universit\"at M\"unchen, D-85748 Garching, Germany
\\
E-mail: \email{waltmann@ph.tum.de}
}

\abstract{
We report on an extensive analysis of FCNC and CPV effects in SUSY theories. We present results for $\Delta F=2$ and $\Delta F=1$ processes governed by $b \to s$ transitions both in the low and high $\tan\beta$ regime, focussing in particular on $S_{\psi\phi}$, the phase of $B_s$ mixing. We emphasize that while the MFV framework is not suited to explain potentially large effects in $S_{\psi\phi}$ as indicated by recent data from CDF and D0, models with large right-right mass insertions in the 32 sector provide natural frameworks to account for such effects. 
Exemplarily we consider two SUSY models based on an abelian and a non-abelian flavor symmetry that show representative flavor structures in the soft SUSY breaking terms and stress that the characteristic correlations among the considered observables allow to distinguish between the different models.
}

\FullConference{
European Physical Society Europhysics Conference on High Energy Physics
\\
July 16-22, 2009
\\
Krakow, Poland
}

\begin{document}

\section{Introduction}
%
\noindent The mixing induced CP asymmetry in $B_s \to \psi\phi$, $S_{\psi\phi}$, is predicted to be tiny in the Standard Model (SM). Analyses of the data on the $B_s \to \psi\phi$ decay taken by CDF and D0 instead find large values for $S_{\psi\phi}$, indicating a tension with the SM number $S_{\psi\phi}^{\rm SM} = \sin2|\beta_s| \simeq 0.036$ at the level of $(2-3)\sigma$~\cite{Barberio:2008fa}
\begin{equation}
S_{\psi\phi} = \sin\left( {\rm Arg}\left( M_{12}^s \right) \right) = \sin\left( 2|\beta_s| - 2\phi_{B_s} \right) \simeq 0.81^{+ 0.12}_{-0.32}~,
\end{equation}
with 
$M_{12}^s = \langle B_s |\mathcal{H}_{\rm eff}| \bar B_s \rangle = C_{B_s} e^{2 i \phi_{B_s}}(M_{12}^s)_{\rm SM}$ and $\phi_{B_s}$ is a New Physics (NP) phase in $B_s$ mixing.

The most popular NP framework where the above mentioned tension can be accounted for is the Minimal Supersymmetric Standard Model (MSSM) where complex contributions to the $B_s$ mixing amplitude $M_{12}^s$ can be naturally generated.
Here we report on a comparative study~\cite{ABGPS} of flavor and CP violating effects in several MSSM scenarios, focusing on the predictions for $S_{\psi\phi}$.

\section{Phenomenology of CP Violation in a Flavor Blind MSSM}
%
\noindent A rather minimalistic supersymmetric (SUSY) framework is a flavor blind MSSM (FBMSSM), where the CKM matrix remains the only source of flavor violation but new CP violating phases are introduced in the soft SUSY breaking sector. 

As shown in~\cite{ABP}, such a scenario can lead to sizeable and highly correlated NP effects in observables sensitive to CP violation in $\Delta F=0$ transitions, as the Electric Dipole Moments (EDMs) of the electron and neutron, and to CP violation in $\Delta F=1$ dipole transitions as the direct CP asymmetry in $b\to s\gamma$ or the mixing induced CP asymmetries in $B \to \phi K_S$ and $B \to \eta^\prime K_S$.
On the other hand it was found that the leading NP contributions to the $\Delta F=2$ mixing amplitudes are not sensitive to the new phases of the FBMSSM, resulting in a SM like $S_{\psi\phi} \simeq 0.03 - 0.05$.

\section{$\mathbf{S_{\psi\phi}}$ in a General MSSM}
%
\noindent In the general MSSM, complex flavor off-diagonal entries in the squark masses, conveniently parameterized in terms of mass insertions (MIs), lead to flavor and CP violating gluino-squark-quark interactions that in turn typically lead to the dominant NP contributions to FCNC processes. 

\begin{figure}[t]\centering
\includegraphics[width=0.29\textwidth]{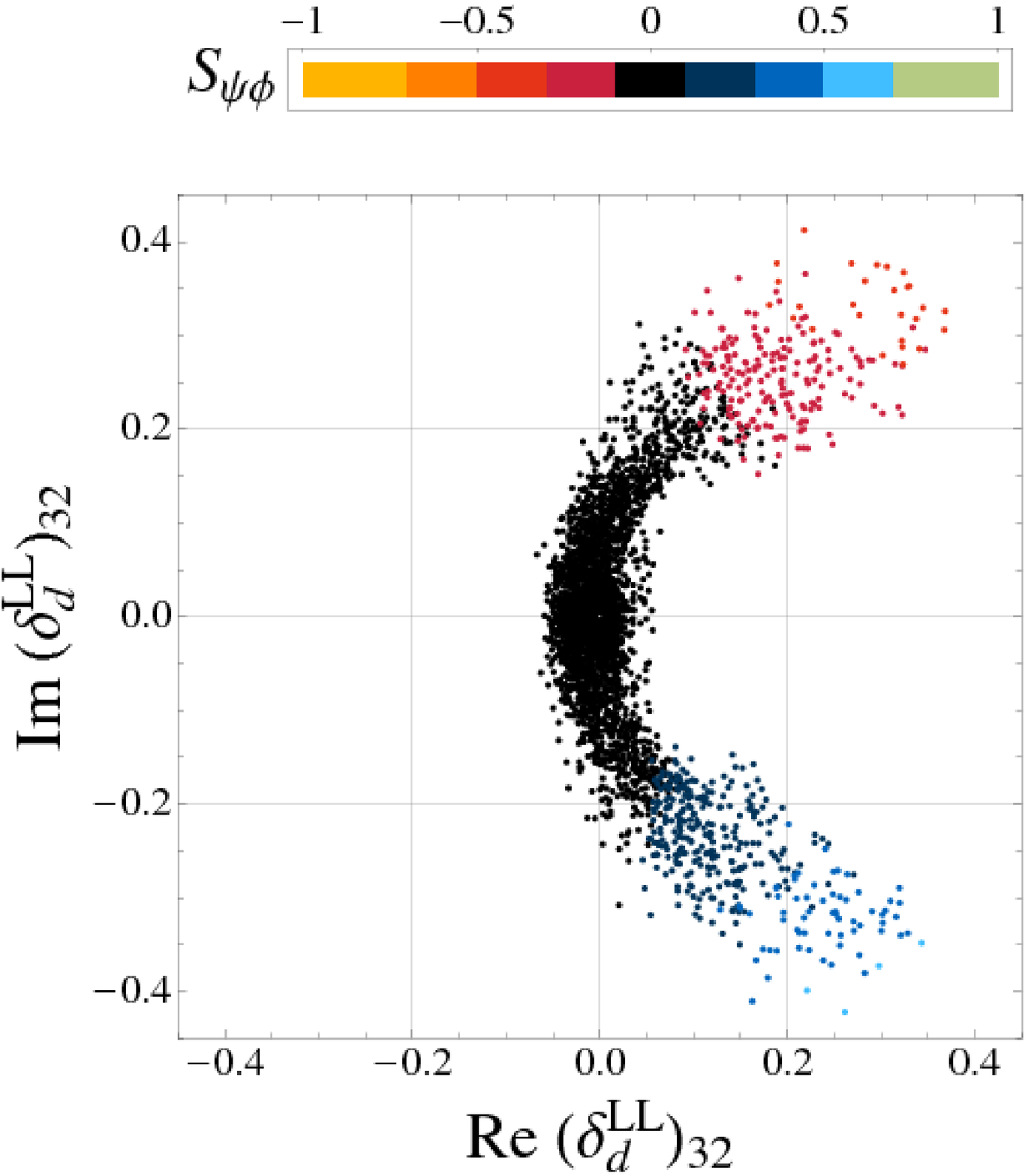}~~~~
\includegraphics[width=0.29\textwidth]{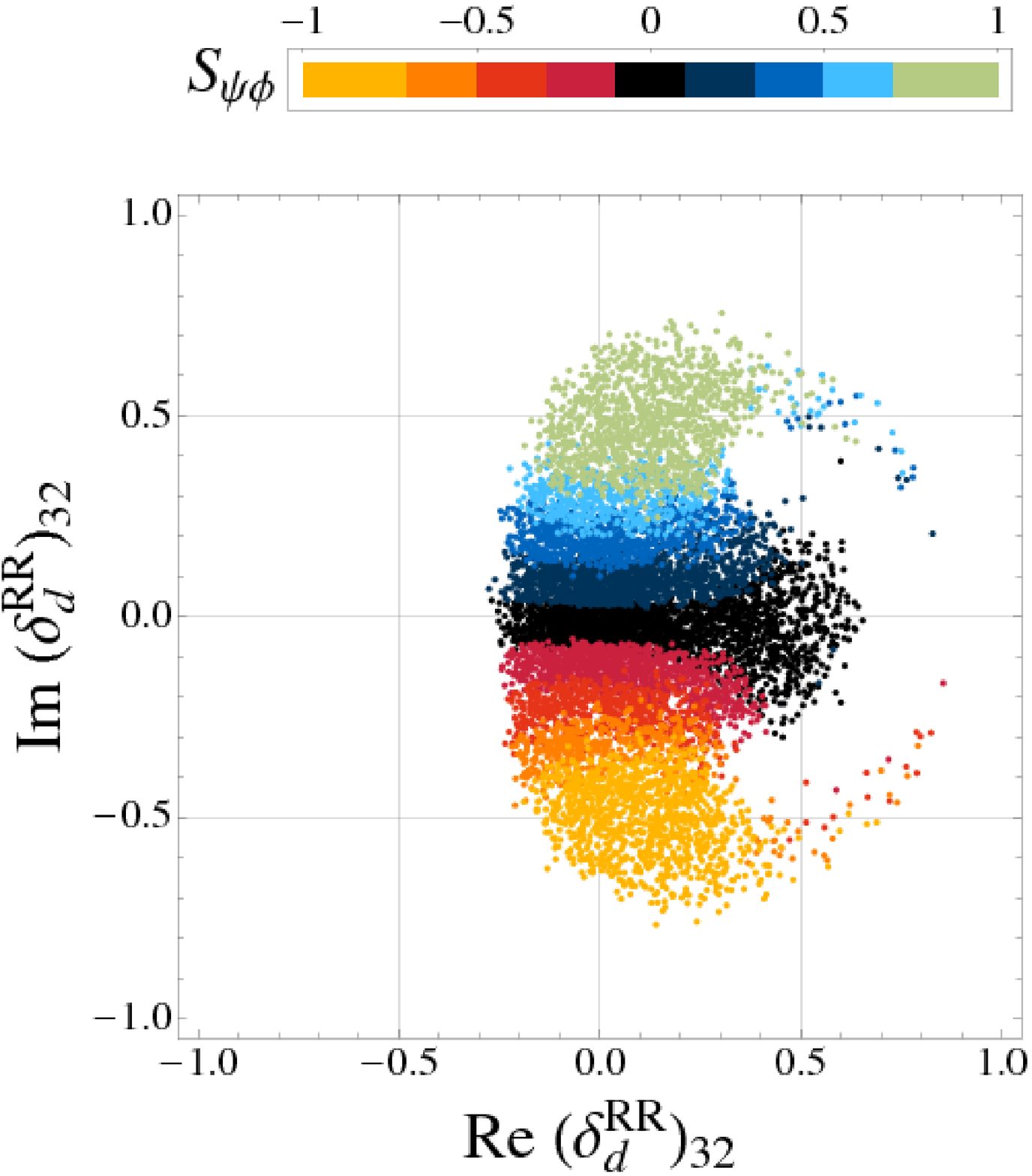}~~~~
\includegraphics[width=0.295\textwidth]{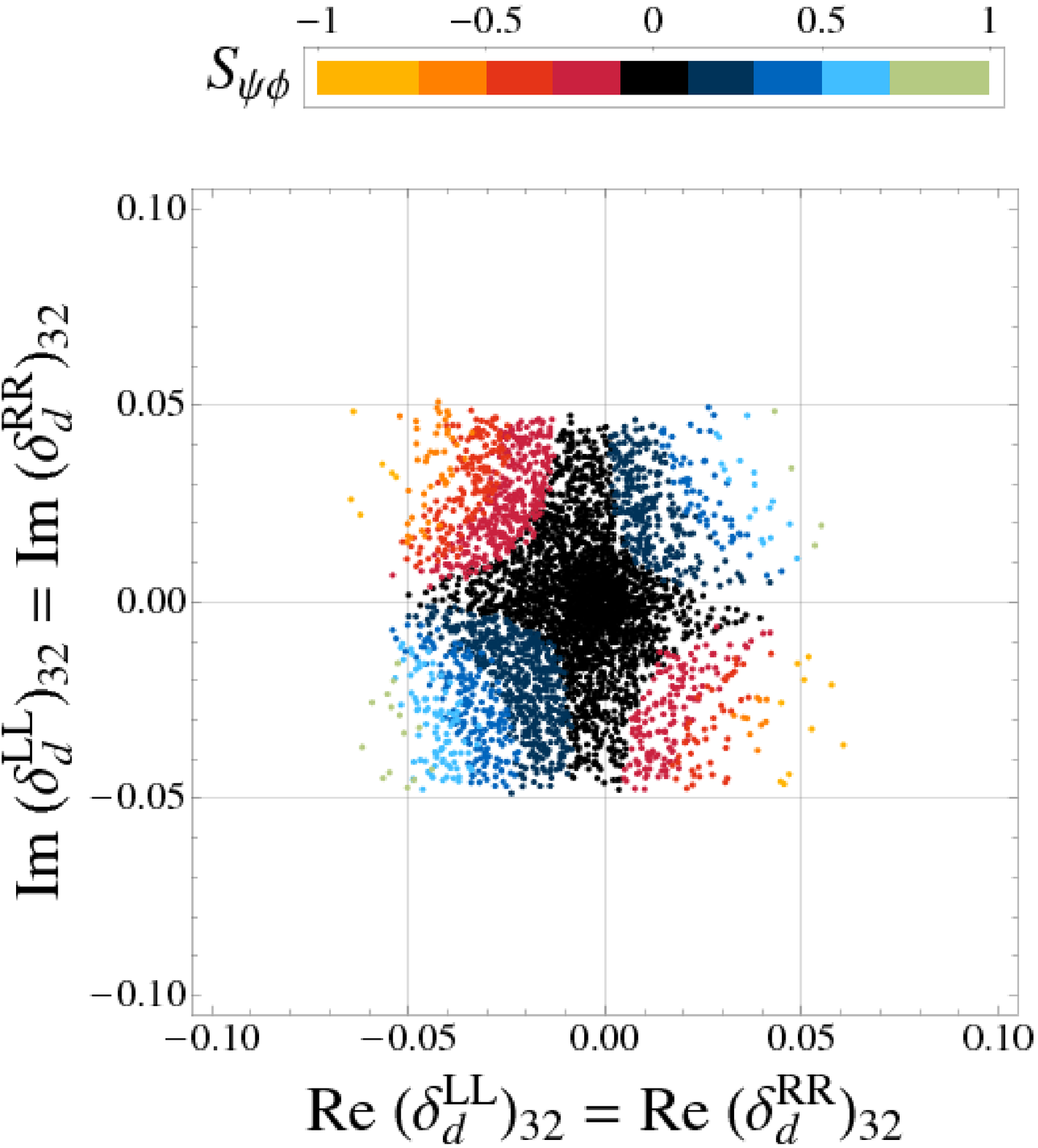}
\caption{\small
Bounds on various MIs $(\delta^{AB}_d)_{32}$ as obtained by imposing in particular the
experimental constraints from BR$(b \to s\gamma)$. BR$(b \to s \ell^+ \ell^-)$ and $\Delta M_s$. The different colors indicate the resulting values for $S_{\psi\phi}$.}
\label{fig:deltas}
\end{figure}
%
In fig.~\ref{fig:deltas} we show the possible values for $S_{\psi\phi}$ arising when {\it i)} only the MI $\left( \delta_d^{LL} \right)_{32}$ (left), {\it ii)} only $\left( \delta_d^{RR} \right)_{32}$ (middle) or {\it iii)} $\left( \delta_d^{LL} \right)_{32} = \left( \delta_d^{RR} \right)_{32}$ (right) are switched on at the low scale, assuming a MSUGRA spectrum with $m_0 < 300~{\rm GeV}$, $M_{1/2} < 200~{\rm GeV}$, $-3m_0 < A_0 < 3 m_0$, $5 < \tan\beta < 15$, $\mu > 0$ and imposing all relevant constraints, in particular from FCNC observables, like BR$(b \to s\gamma)$, BR$(b \to s \ell^+ \ell^-)$ and $\Delta M_s$. 
On the other hand we do not show the cases of $\left( \delta_d^{LR} \right)_{32}$ and $\left( \delta_d^{RL} \right)_{32}$ MIs. There the constraint from BR$(b \to s\gamma)$ is so strong that effects in $S_{\psi\phi}$ are hardly possible.

We observe that for the above choice of parameters especially in the case {\it ii)} huge values for the $B_s$ mixing phase in the whole range $-1 < S_{\psi\phi} < 1$ are possible. Also in the case {\it iii)} one gets large effects in $S_{\psi\phi}$ even for small, CKM like MIs $\left( \delta_d^{LL} \right)_{32} = \left( \delta_d^{RR} \right)_{32} \simeq 0.04$.
We remark that a $\left( \delta_d^{LL} \right)_{32} \propto V_{ts}^* V_{tb}$ is always induced radiatively in the running from the high (GUT) scale down to the electroweak scale. Thus in both cases {\it ii)} and {\it iii)} large contributions to the mixing amplitude come from left-right operators generated by gluino boxes, that are strongly enhanced through renormalization group effects~\cite{Ciuchini:1997bw} and through a large loop function~\cite{Gabbiani:1996hi}.

In conclusion, scenarios with sizeable right-handed currents, induced by $\left( \delta_d^{RR} \right)_{32}$ MIs, provide natural frameworks where large effects in the $B_s$ mixing amplitude can be generated and non-standard values for $S_{\psi\phi}$ can be expected.

\section{Predictions for $\mathbf{S_{\psi\phi}}$ in SUSY Flavor Models}

\noindent Large $\left( \delta_d^{RR} \right)_{32}$ MIs can naturally arise e.g. in the context of SUSY GUTs~\cite{Moroi:2000mr} and within SUSY flavor models. In the following we concentrate on two examples: an abelian flavor model by Agashe, Carone (AC)~\cite{Agashe:2003rj} and a non-abelian $SU(3)$ flavor model by Ross, Velasco-Sevilla, Vives (RVV2)~\cite{Ross:2004qn}. 
The AC model is based on a $U(1)$ flavor symmetry and a non-trivial higher-dimensional topography and suppresses FCNCs in the $b \to d$ and $s \to d$ sectors through the alignment mechanism~\cite{Nir:1993mx}. As all models based on abelian flavor symmetries, it predicts $\left( \delta_u^{LL} \right)_{21} = O(\lambda_C)$, leading to large effects in $D_0$ - $\bar D_0$ mixing. In addition it predicts $\left( \delta_d^{RR} \right)_{32} = O(1)$ such that large NP contributions to $b \to s$ transitions are expected.
The RVV2 model on the other hand, predicts almost degenerate squarks of the first two generations which implies tiny contributions to $D_0$ - $\bar D_0$ mixing. However, effects in $K$ mixing and especially in $\epsilon_K$ can still be large. Similar to the AC model it predicts a sizeable $\left( \delta_d^{RR} \right)_{32} = O(0.15)$.

\begin{figure}[t]\centering
\includegraphics[width=0.23\textwidth]{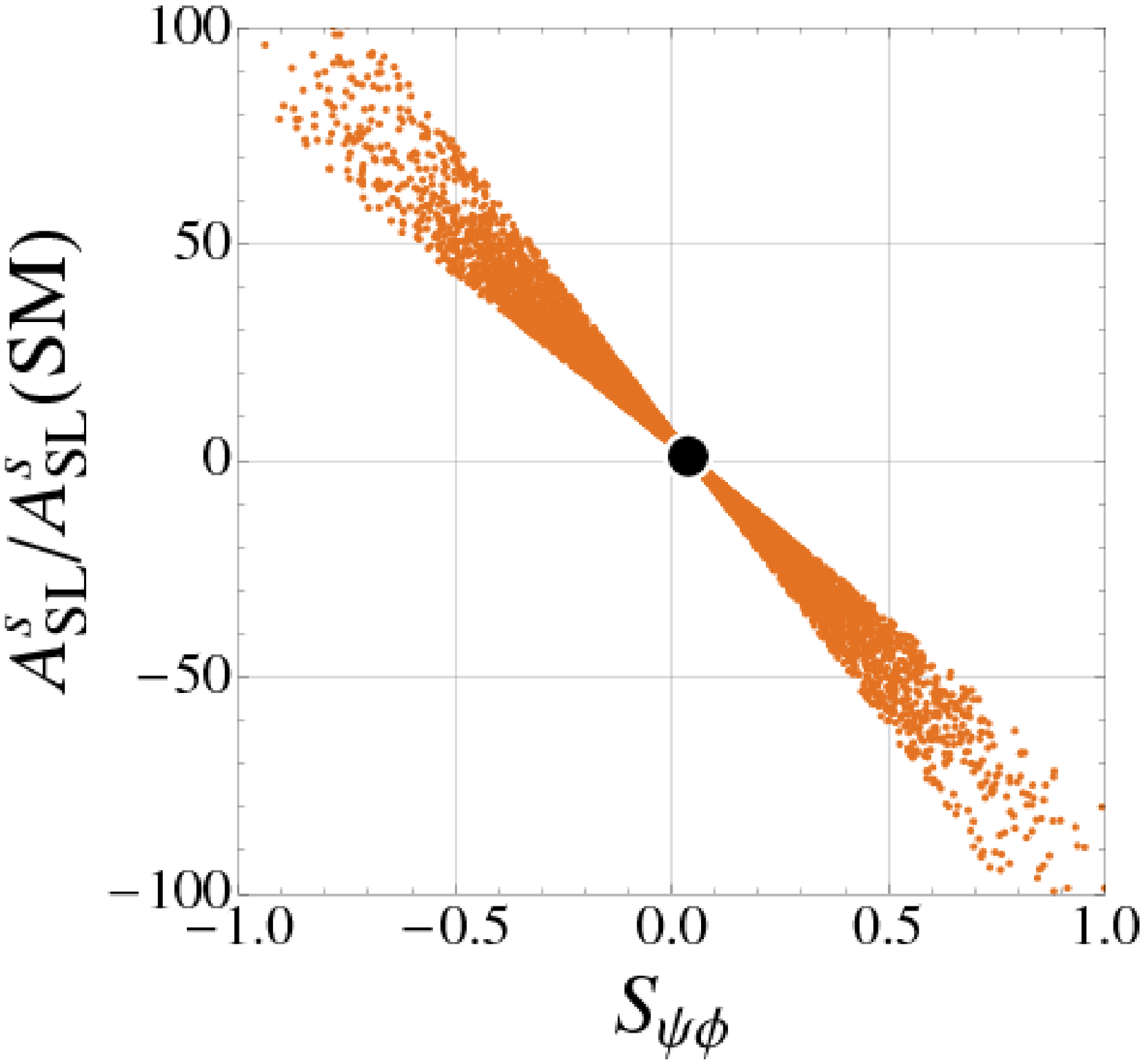}~
\includegraphics[width=0.245\textwidth]{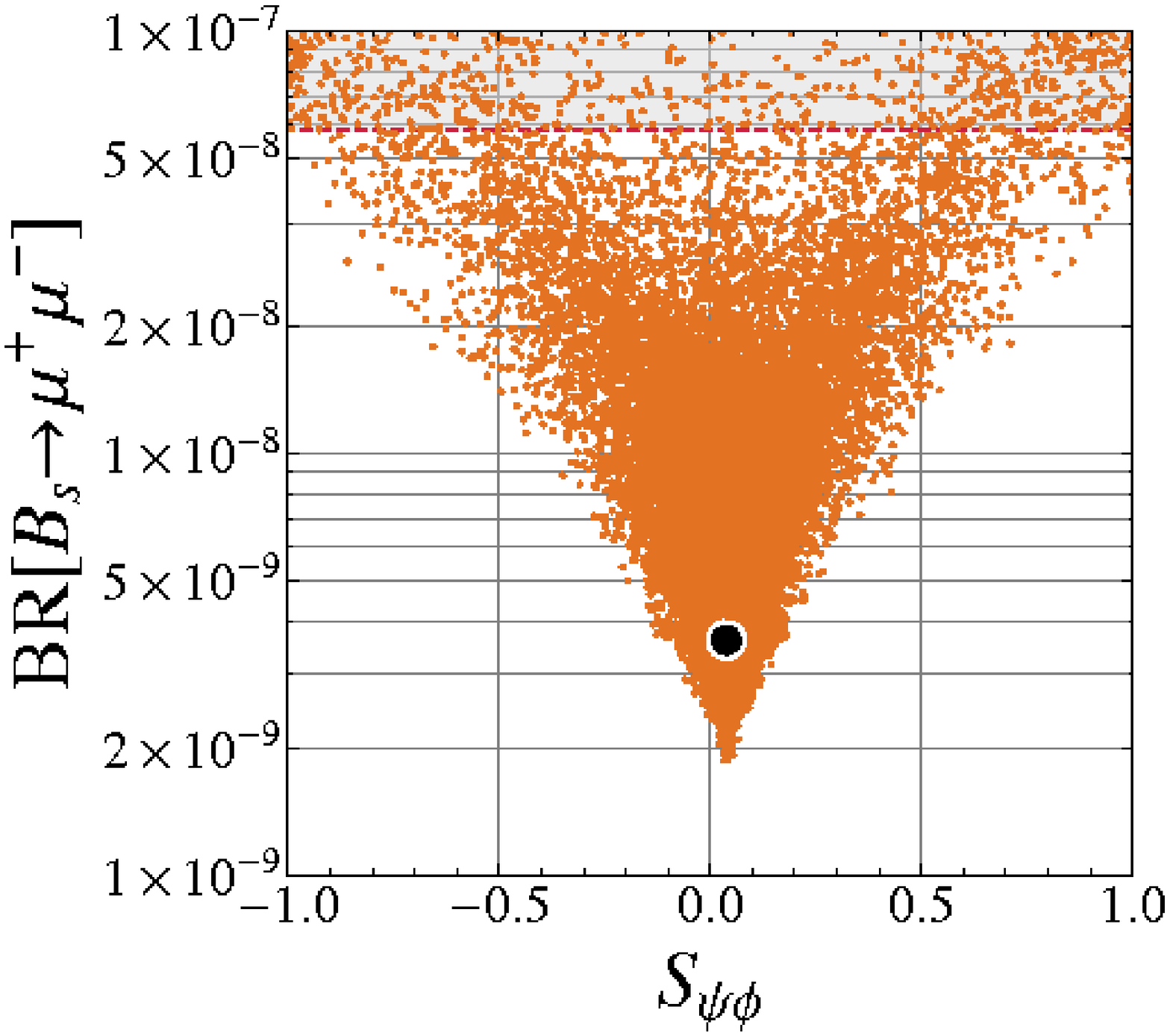}~~~~~
\includegraphics[width=0.23\textwidth]{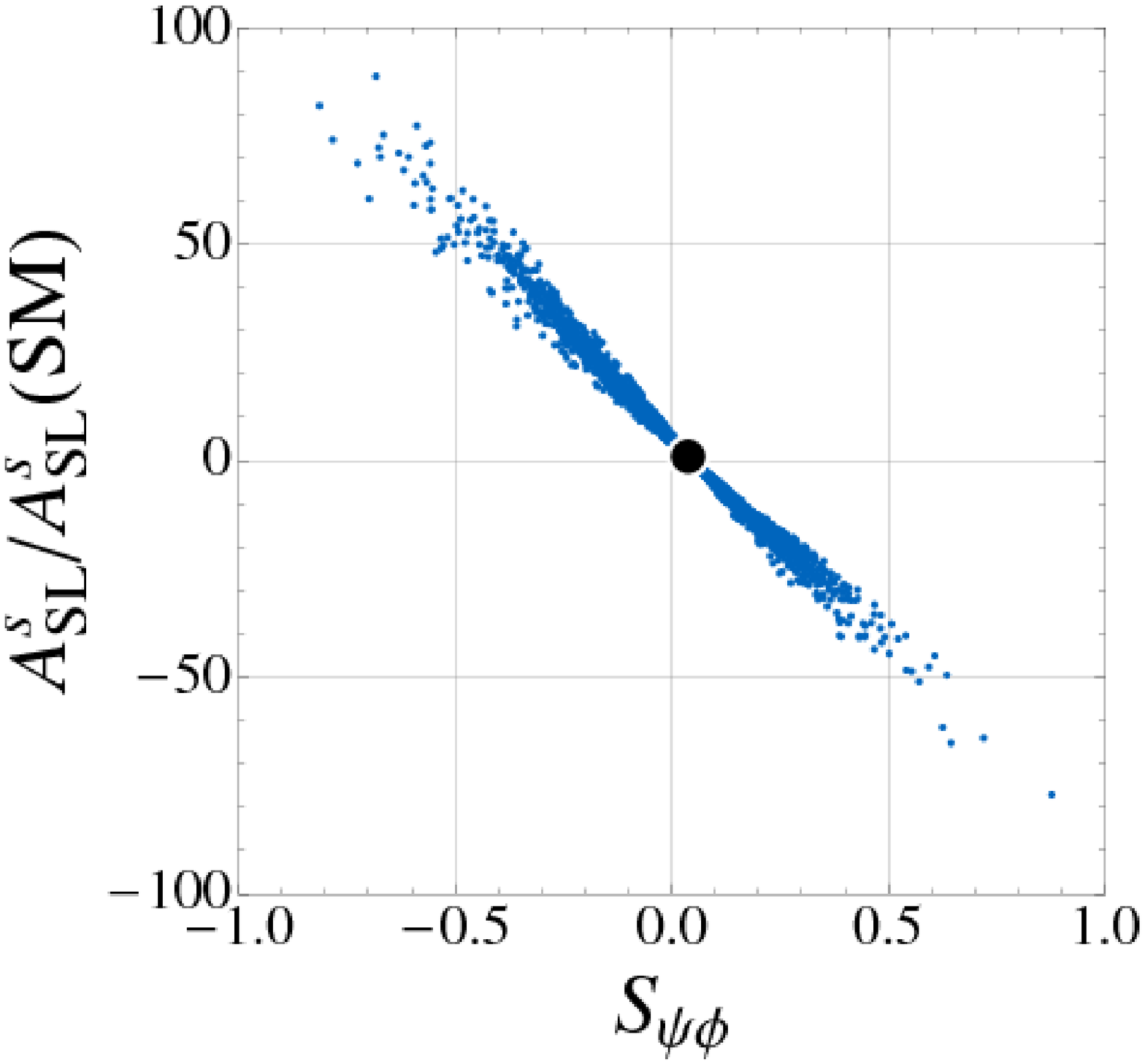}~
\includegraphics[width=0.245\textwidth]{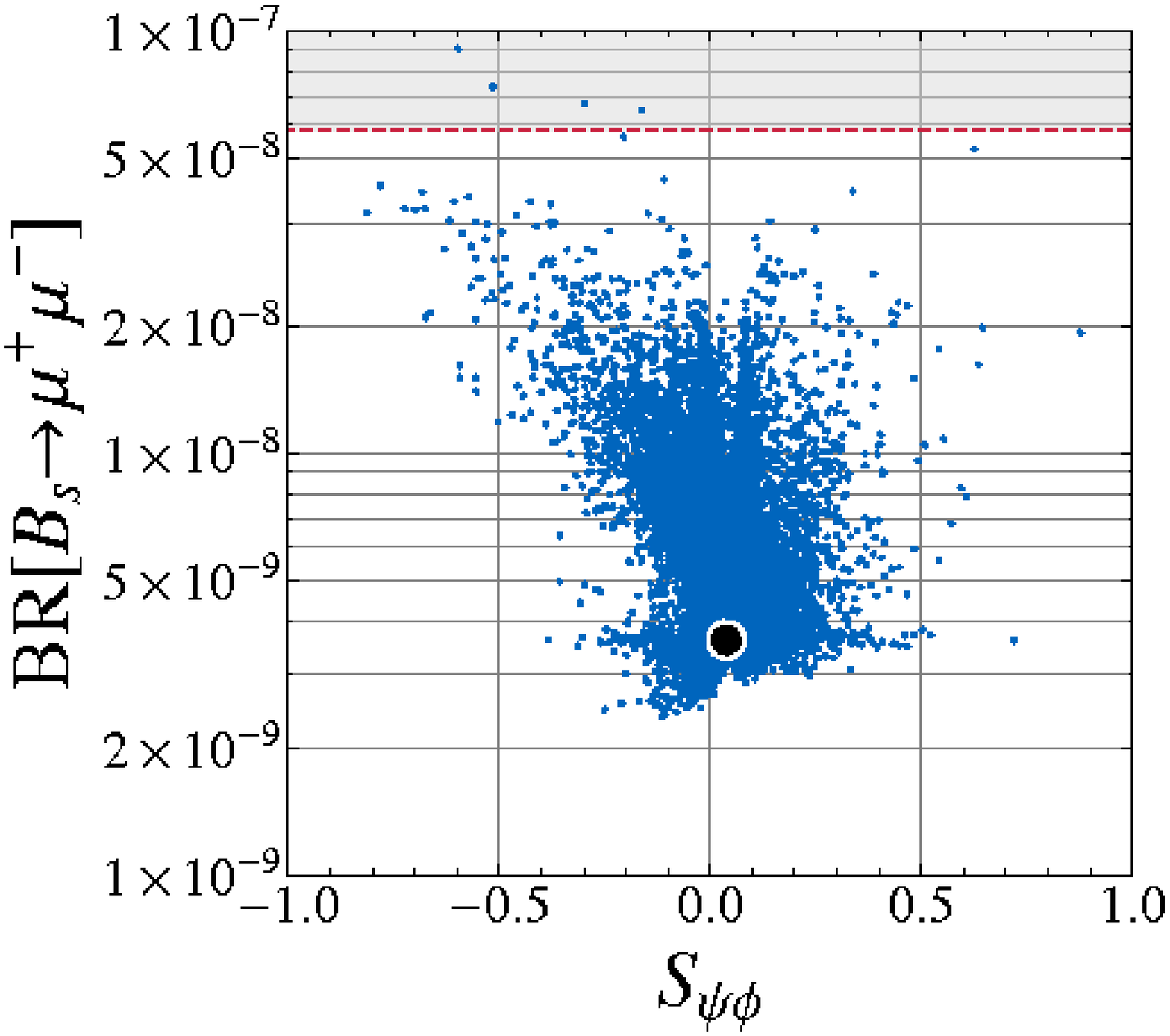}
\caption{\small
Correlations of $S_{\psi\phi}$ with $A_{\rm SL}^s$ and BR$(B_s \to \mu^+\mu^-)$ in the abelian flavor model of~\cite{Agashe:2003rj} (left) and the non-abelian flavor model of~\cite{Ross:2004qn} (right).}
\label{fig:flavor_models}
\end{figure}

In fig.~\ref{fig:flavor_models} we show the predictions of the two models for $S_{\psi\phi}$ and its correlations with the semileptonic asymmetry $A_{\rm SL}^s$~\cite{Ligeti:2006pm} and BR$(B_s \to \mu^+\mu^-)$, based on a MSUGRA spectrum with $m_0 < 2~{\rm TeV}$, $M_{1/2} < 1~{\rm TeV}$, $-3m_0 < A_0 < 3 m_0$, $5 < \tan\beta < 55$ and $\mu > 0$. In the AC (RVV2) model, values for $S_{\psi\phi}$ in the range $-1 < S_{\psi\phi}^{\rm AC} < 1$ ($-0.7 < S_{\psi\phi}^{\rm RVV2} < 0.7$) are possible, leading to an enhancement of $A_{\rm SL}^s$ of up to two orders of magnitude compared to its SM prediction.

In the AC model there exists a strong correlation of $S_{\psi\phi}$ with BR$(B_s \to \mu^+\mu^-)$ implying a lower bound of BR$(B_s \to \mu^+\mu^-) \gtrsim 10^{-8}$ for $|S_{\psi\phi}| \gtrsim 0.3$. This correlation shows, that in the AC model the large values of $S_{\psi\phi}$ are due to double Higgs penguin contributions that, in presence of $\left( \delta_d^{RR} \right)_{32}$, are not suppressed by $m_s/m_b$ as it happens instead in minimal flavor violating frameworks. Box contributions play only a minor role in the AC model, once the experimental constraints from $D_0$ - $\bar D_0$ mixing are imposed.

In the RVV2 model the correlation between $S_{\psi\phi}$ and BR$(B_s \to \mu^+\mu^-)$ is lost to a large extent, due to the absence of the $D_0$ - $\bar D_0$ mixing constraint and the richer flavor structure of the model.

\section{Conclusions and Outlook}

\noindent The MSSM is a very well motivated extension of the SM able to address the recently reported non-standard effects in $S_{\psi\phi}$. While $S_{\psi\phi}$ remains SM like in scenarios where the CKM matrix is the only source of flavor violation, large CP violation in $B_s$ mixing is naturally predicted in models with sizeable right-handed currents, induced by complex $\left( \delta_d^{RR} \right)_{32}$ mass insertions.

The extensive analysis of~\cite{ABGPS} confirms this statement also in the context of concrete flavor models and outlines possible strategies to distinguish between the different models by means of a comparative study of processes governed by $b \to s$ transitions and their correlations with processes governed by $b \to d$ transitions, $s \to d$ transitions, $D_0$ - $\bar D_0$ oscillations, lepton flavor violating decays, EDMs and $(g-2)_\mu$.




\begin{thebibliography}{99}

\bibitem{Barberio:2008fa}
  E.~Barberio {\it et al.}  [Heavy Flavor Averaging Group],
  arXiv:0808.1297 [hep-ex].

\bibitem{ABGPS}
  W.~Altmannshofer, A.~J.~Buras, S.~Gori, P.~Paradisi and D.~M.~Straub,
  arXiv:0909.1333 [hep-ph].


\bibitem{ABP}
  W.~Altmannshofer, A.~J.~Buras and P.~Paradisi,
  Phys.\ Lett.\  B {\bf 669} (2008) 239
  [arXiv:0808.0707 [hep-ph]].

\bibitem{Ciuchini:1997bw}
  M.~Ciuchini, E.~Franco, V.~Lubicz, G.~Martinelli, I.~Scimemi and L.~Silvestrini,
  Nucl.\ Phys.\  B {\bf 523} (1998) 501
  [arXiv:hep-ph/9711402],
  A.~J.~Buras, M.~Misiak and J.~Urban,
  Nucl.\ Phys.\  B {\bf 586} (2000) 397
  [arXiv:hep-ph/0005183].

\bibitem{Gabbiani:1996hi}
  F.~Gabbiani, E.~Gabrielli, A.~Masiero and L.~Silvestrini,
  Nucl.\ Phys.\  B {\bf 477} (1996) 321
  [arXiv:hep-ph/9604387].

\bibitem{Moroi:2000mr}
  T.~Moroi,
  JHEP {\bf 0003} (2000) 019
  [arXiv:hep-ph/0002208],
  D.~Chang, A.~Masiero and H.~Murayama,
  Phys.\ Rev.\  D {\bf 67} (2003) 075013
  [arXiv:hep-ph/0205111]
  B.~Dutta and Y.~Mimura,
  Phys.\ Lett.\  B {\bf 677} (2009) 164
  [arXiv:0902.0016 [hep-ph]],

\bibitem{Agashe:2003rj}
  K.~Agashe and C.~D.~Carone,
  Phys.\ Rev.\  D {\bf 68} (2003) 035017
  [arXiv:hep-ph/0304229].

\bibitem{Ross:2004qn}
  G.~G.~Ross, L.~Velasco-Sevilla and O.~Vives,
  Nucl.\ Phys.\  B {\bf 692} (2004) 50
  [arXiv:hep-ph/0401064],
  L.~Calibbi, J.~Jones-Perez and O.~Vives,
  Phys.\ Rev.\  D {\bf 78} (2008) 075007
  [arXiv:0804.4620 [hep-ph]],
  L.~Calibbi, J.~Jones-Perez, A.~Masiero, J.~h.~Park, W.~Porod and O.~Vives,
  arXiv:0907.4069 [hep-ph].

\bibitem{Nir:1993mx}
  Y.~Nir and N.~Seiberg,
  Phys.\ Lett.\  B {\bf 309} (1993) 337
  [arXiv:hep-ph/9304307].

\bibitem{Ligeti:2006pm}
  Z.~Ligeti, M.~Papucci and G.~Perez,
  Phys.\ Rev.\ Lett.\  {\bf 97} (2006) 101801
  [arXiv:hep-ph/0604112].

\end{thebibliography}
\end{document}